# THz ultrastrong light-matter coupling
Highly scalable semiconductor-based platform for on-chip probing of strong coupling physics

**Giacomo Scalari,[a] Curdin Maissen,[a] Sara Cibella,[b] Roberto Leoni[b], Christian Reichl[c], Werner Wegscheider[c], Mattias Beck[a], Jérôme Faist[a]**

[a] Institute of Quantum Electronics, Eidgenössische Technische Hochschule Zürich, Switzerland
[b] Istituto di Fotonica e Nanotecnologie (IFN), CNR, via Cineto Romano 42, 00156 Rome, Italy
[c] Laboratory for Solid State Physics, Eidgenössische Technische Hochschule Zürich, Switzerland

**Abstract**. Cavity photon resonators with ultrastrong light-matter interactions are attracting interest both in semiconductor and superconducting systems displaying the capability to manipulate the cavity quantum electrodynamic ground state with controllable physical properties. Here we review a series of experiments aimed at probing the ultrastrong light-matter coupling regime, where the vacuum Rabi splitting Ω is comparable to the bare transition frequency $\omega_c$. We present a new platform where the inter-Landau level transition of a two-dimensional electron gas (2DEG) is strongly coupled to the fundamental mode of deeply subwavelength split-ring resonators operating in the mm-wave range. Record-high values of the normalized light-matter coupling ratio $\frac{\Omega}{\omega_c} = 0.89$ are reached and the system appears highly scalable far into the microwave range.

**Address all correspondence to:** Giacomo Scalari, scalari@phys.ethz.ch

## 1 Introduction: hybrid metasurface for THz light-matter coupling.

Enhancement and tunability of light-matter interaction is crucial for fundamental studies of cavity quantum electrodynamics (QED) and for applications in classical and quantum devices [1]. The coupling between one cavity photon and one elementary electronic excitation is quantified by the vacuum Rabi frequency Ω [1]. The non-perturbative strong light-matter coupling regime is achieved when Ω is larger than the loss rates of the cavity and electronic excitations. Recently, growing interest has been generated by the so-called ultrastrong coupling regime [2,3] which is obtained when the vacuum Rabi frequency Ω becomes an appreciable fraction of the unperturbed frequency $\omega_c$. In such a regime the system under consideration has to be treated beyond the rotating wave approximation, analogously to what happens in spin resonance for high irradiation powers leading to effects such as the Bloch-Siegert shift. The consequence of these additional terms is the modification of the ground and excited state properties of the light-matter coupled system. The ultrastrong coupling regime of cavity QED has been predicted to display intriguing and peculiar quantum electrodynamics features: Casimir-like squeezed vacuum photons upon either non-adiabatic change or periodic modulation in the coupling energy [3], non-classical radiation from chaotic sources [4] and others.
Semiconductor-based systems operating in the Mid-IR and THz are especially attractive for the study of this peculiar regime as very large intersubband dipole moments $d$ can be achieved [5,6,7]. The system can also benefit from the $\sqrt{N}$ enhancement of the light-matter coupling by deriving from the simultaneous coupling of N electronic excitations with dipole $d$ to the vacuum fluctuations $E_{vac}$ of the same cavity mode. We recall that the r.m.s. vacuum field fluctuations at an angular frequency $\omega_c$ are related to the cavity volume as $E_{vac} \sim \sqrt{\frac{\hbar \omega_c}{\epsilon V_{cav}}}$ where $V_{cav}$ is the cavity mode volume and $\epsilon = n_{eff}^2$ where $n_{eff}$ is the effective refractive index [8]. In our vacuum field Rabi splitting experiments the average number of photons in the cavity is usually very small and the normal mode splitting is only due to the vacuum cavity field.

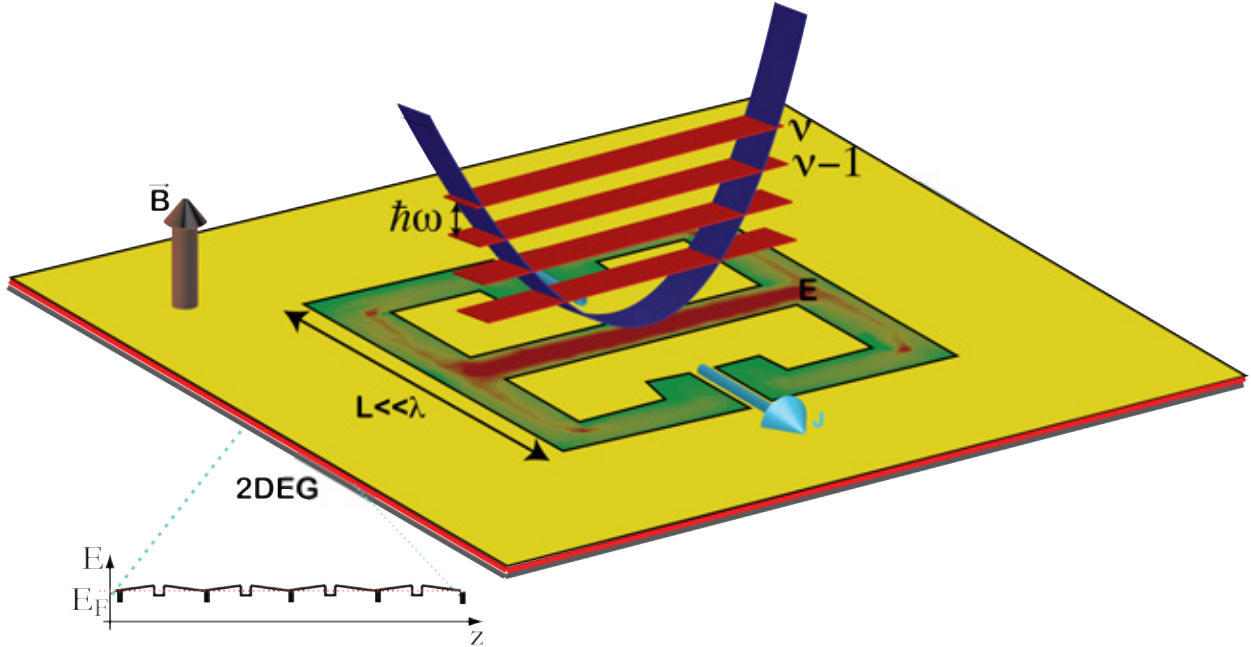

**Fig. 1** *Schematic of the light-matter coupling experiments with metallic metasurfaces and 2DEG. The magnetic field applied perpendicularly to the surface of the 2DEG gives rise to a parabolic potential where equidistantly spaced Landau levels are formed. The complementary metasurface is based on split-ring resonators which are LC circuits where the current J and the electric field E are separated spatially. The in-plane electric field E that couples with the TE polarized Inter-Landau-level transition (calculated with 3D finite element solver) is plot in color scale and is mainly concentrated in the gap between the metals. Oscillating currents J are flowing in the metallic apertures.*

The matter part of our system is still constituted by a semiconductor quantum well but we do not exploit intersubband transitions coming from size quantization. The THz and microwave-active transition we consider is the one arising between consecutive Landau levels which are created when the 2-dimensional electron gas (2DEG) is immersed in a DC magnetic field parallel to the growth axis of the heterostructure. As discussed in the introduction of Ref. [9], starting from the light-matter coupling relation $\hbar\Omega = d \times E_{vac}$ it can be shown that the normalized light-matter coupling ratio for this cyclotron-based system scales in the following way with the relevant physical parameters [10,11]:

$$\frac{\Omega}{\omega_c} = \sqrt{\frac{\lambda}{V_{cav}} \int_{2DEG} A ds \frac{\alpha \nu}{\epsilon}}$$

Here $\nu = \frac{2\pi\hbar\rho}{eB}$ is the 2DEG fill factor, $\rho$ is the 2DEG sheet carrier density, $\alpha$ is the fine structure constant, e the elementary charge and A is a quantity related to the vector potential of the THz mode. This equation highlights the dependence on $\nu$ as well as on the ratio between the wavelength and the cavity dimensions. Differently from Refs. [9,12], we experimentally implement this scheme using a strongly subwavelength resonant cavity constituted by split-ring resonators which are the building blocks of metamaterials and metasurfaces [13] [14]. One of the salient features of the split-ring resonators is their extreme sub-wavelength dimensions which allow the concentration of the electromagnetic fields in extremely reduced volumes. For our purposes this greatly enhances the vacuum field fluctuations allowing us to achieve extremely high normalized coupling ratios $\frac{\Omega}{\omega_c}$. A schematic of the typical sample employed in our experiments is presented in Fig. 1. We deposit the metallic metasurface (Ti/Au) directly on top of the semiconductor heterostructure containing the 2DEG. The electric field is mainly concentrated in the capacitor element of the LC circuit, which constitutes the split-ring resonator. The fringing field penetrates for a couple of micrometers in the semiconductor providing the electric field at the location of the quantum well containing electrons.

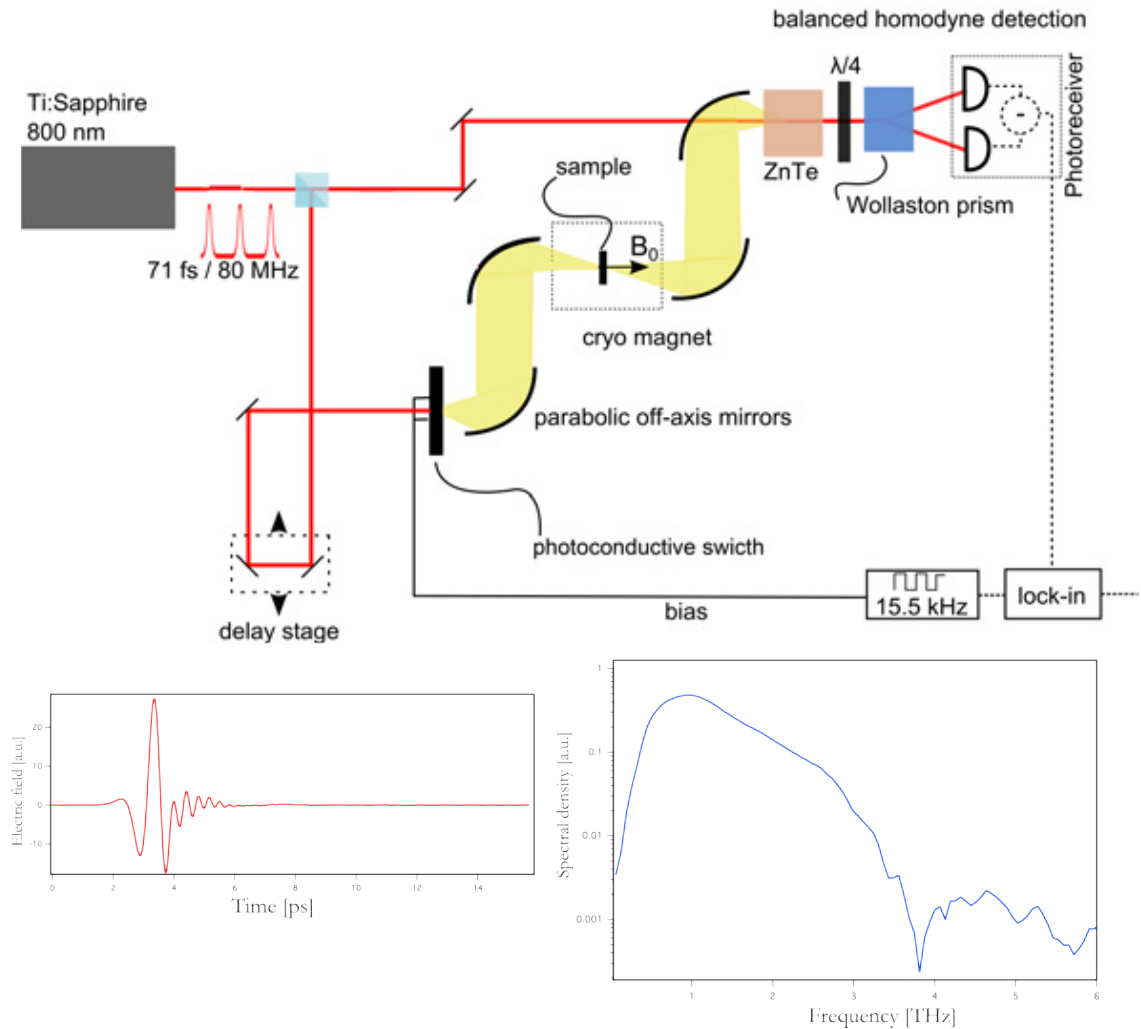

**Fig. 2**: *experimental setup used for the transmission experiments presented in this paper. The ultrafast Ti:sapphire laser of pulse width <75 fs illuminates a photoconductive switch producing a broadband THz pulse of length 2 ps The THz pulse propagates parallel to the DC magnetic field direction in order couple to the cyclotron transition of the 2DEG. THz-TDS allows coherent detection of the amplitude and phase of the THz electric field [15]. In the lower panel is visible the THz electric field pulse as a function of time (red trace) and its Fourier transform which extends*

*above 3 THz.*

The samples are investigated using a transmission spectrometer based on a THz-Time Domain Spectroscopy (TDS) setup coupled to a liquid helium cryostat equipped with a superconducting magnet (maximum field of 11 T) and a variable temperature insert (1.9-300 K). The scheme of the setup with the essential components is displayed in Fig. 2.

Examining the data of Fig.3(a), at zero magnetic field we observe two resonances whose origin is qualitatively different: the lowest frequency mode at about 900 GHz is attributed to the LC resonance, where counterpropagating currents circulate in the inductive part and the electric field is enhanced mainly in the capacitor gap [14] (see schematic in Fig. 1) . The second mode at approximately 2.3 THz is attributed to the "cut wire" behavior where a λ/2 (or dipolar) resonance is excited along the sides of the metaparticle. These values correspond well to simulations done employing 3D finite element modeling (CST microwave studio package) [16]. The presence of conductive layers underneath alters the frequency and the quality factor of these resonances. In the data reported in Fig.3(a) we observe the evolution of the sample transmission as a function of the applied magnetic field (normalized to the electric field of the reference 2DEG wafer at B=0). An heterostructure with $n_{QW}$=4 and $\rho$=4.5 x $10^{11} cm^{-2}$ is used as an active medium. As the magnetic field is swept a profound modification of the sample transmission is observed. The possibility to tune in a continuous way the material excitation allows to follow the evolution of polaritonic states as the system is driven from the uncoupled regime to the strongly coupled one. We observe two successive anticrossings when the cyclotron energy matches the first and the second resonator modes [16]. In Fig. 3(b) we extract the positions of the minima of sample transmission and plot the dispersion curves for the polariton eigenvalues as a function of the magnetic field. The curves are calculated using a full quantum mechanical treatment of the system, obtained generalizing the theory described in Ref.[9] to the case of a zero-dimensional resonator exhibiting two modes with different transverse wavevectors.

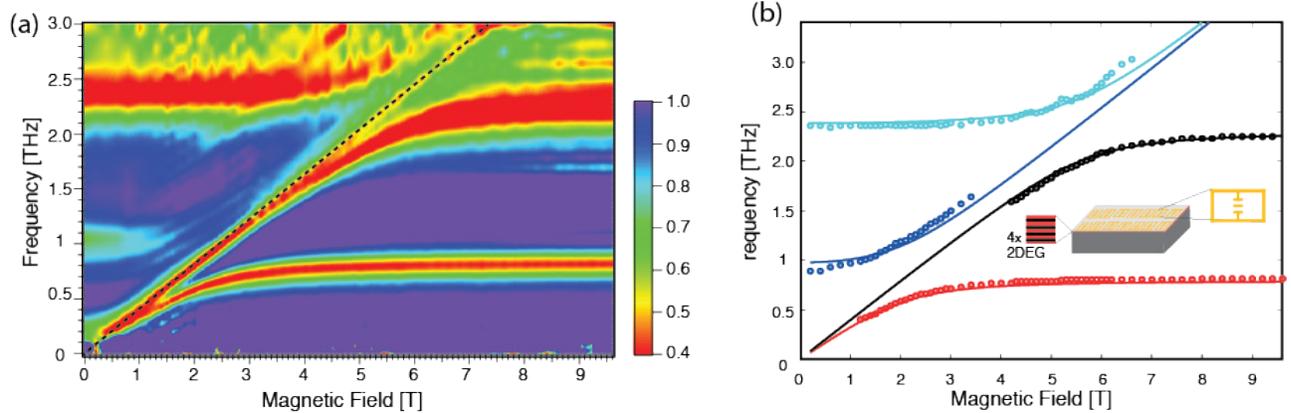

**Fig. 3** *(a): Transmission of a sample as a function of the applied magnetic field. The reference is a plain 2DEG sample at B=0 without resonators on top and the measurement is performed at T=10 K. The black dotted line highlights the cyclotron signal coming from the weakly coupled 2DEG material which is left in-between the resonators. (b): best fit using the theory of Ref.[9] with the extracted transmitted minima positions for the two orthogonal modes of the split ring electronic resonator; the fitting parameter is $\frac{\Omega}{\omega_c}$. Inset: schematic of the sample with the metamaterial and the n=4 2DEG used. Adapted with permission from Ref.[16])*

In order to fit the experimental data, we need to know the resonator modes frequencies as well as the strength of their couplings. For each cavity mode we assumed the asymptotic value of the corresponding lower polariton branch to coincide with the frequency of the unloaded resonator. The coupling strength for the two modes cannot be directly measured, we thus applied a best-fit procedure following the least square method. The minimal error (5%) is obtained for a normalized coupling ratio $\frac{\Omega}{\omega_c} = 0.36$ for the first mode (the LC mode ) and we obtain $\frac{\Omega}{\omega_c} = 0.16$ for the second mode. The second mode presents a lower normalized ratio because of the $\sqrt{\nu}$ dependence and also because the normalized cavity volume is larger than the LC mode case, yielding smaller vacuum fluctuations.

## 2 Ultrastrong coupling with complementary metasurfaces

In the first series of experiments[16][17] presented in the previous paragraph we used metasurfaces of split-ring resonators that, when measured in transmission, display an absorption dip at the resonant frequency of the LC mode. The characteristic anticrossing behavior of the polaritonic system is probed by sweeping the magnetic field. This provides a linear tuning of the cyclotron energy (neglecting magneto-plasmonic contributions at low frequencies). With such experimental arrangement we have observed three main absorption features: two are related to the light-matter coupled system and the third middle peak is the cyclotron signal coming from the material which lies in between the resonators (black dotted line in Fig.3(a)). This cyclotron signal is only weakly coupled to the metasurface and follows the expected linear dispersion for a cyclotron transition. To resolve with higher accuracy the spectroscopic features of the polaritonic branches, we changed the configuration of the metamaterial cavity by employing a complementary cavity. As already shown in Ref. [18] and other papers, the complementary metamaterial is obtained by exchanging the roles of the vacuum areas and of the metals composing the metasurface. The resulting metasurface is constituted of a metallic sheet with openings with the shapes of the resonators. When excited with an electric field perpendicular to the one used in the case of the direct metamaterial the resonator will display a transmission spectrum, which, according to Babinet's principle, is complementary to the one shown by the split ring resonators (i.e. transmission peaks instead of transmission dips). In Fig. 4(a,b) we report the in-plane electric field distribution for the standard metamaterial and its complementary tuned to resonate at 500 GHz. In the same Figure in panels (c,d) are reported the experimental data obtained with both kinds of cavities fabricated on top of the same n=1 and *ρ=3 x 10$^{11}$cm$^{-2}$* quantum well heterostructure[11]. It is evident the absence of the weakly coupled cyclotron signal in Fig. 4(d) and a very clean anticrossing is observed. We measure normalized coupling ratios of 0.34 and 0.27 for the direct and complementary sample respectively. The only difference in the two samples is the geometry and in particular the gap forming the capacitor. This difference leads to an increased effective volume for the complementary split-ring resonator and a reduced coupling strength The high potential of the split-ring resonators for ultrastrong coupling experiments becomes evident, when comparing the measured coupling strengths to the prediction for a Fabry-Perot resonator for the same filling factor, which, according to Ref.[9] , is 0.15. Both split-ring resonators (direct and complementary) are clearly outperforming the Fabry-Perot microcavity by up to more than a factor 2.

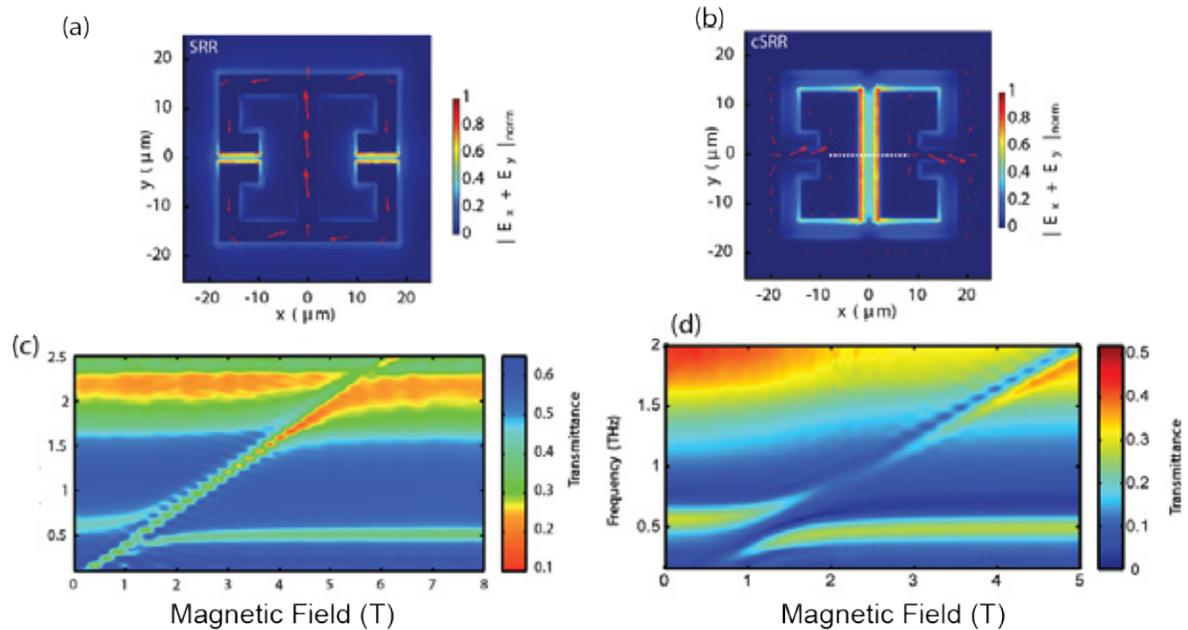

**Fig. 4** *In panels (a) and (b), FEM simulations of the direct (a) and complementary (b) split-ring resonator show the in-plane 100 nm below the semiconductor surface (color scale) and the current distribution in the gold structure (red arrows). Complementary split-ring resonators show a complementary transmission spectrum compared to their direct counterpart. In panel (c) we report the contour plot of the transmittance through the* direct *split-ring resonator on a single 2DEG ($n_{QW}=1$) 2. The anticrossing of the LC mode with the cyclotron transition takes place at Bres =1.2T. Uncoupled areas of the 2DEG give rise to the minimum evolving linearly with B (transmission measurements were performed at an interval of ΔB = 0.2 T; this periodicity appears as an apparent modulation of the transmission). (d) The same measurement for the complementary version of of the split-ring resonator on the same single quantum well (color scale is inverted). The uncoupled parts of the 2DEG do not contribute to the signal Both measurements are performed at T=10 K (adapted with permission from Ref.[11] ).*

### 3 Superconducting metamaterials: higher Q factors and switching capability

The intriguing quantum optical predictions for an ultrastrongly coupled system rely on a non-adiabatic modulation of the system's parameters[3]; in our case on timescales shorter than the inverse of the Rabi frequency of the system (10-2.5 ps). The fabrication of a superconducting cavity offers an interesting opportunity in this direction, since the cavity characteristics strongly depend on the regime of the superconductor. A similar approach led to the experimental demonstration of the dynamical Casimir effect [19]. The presence of the superconductor will also be beneficial since it will increase the Q factor and enhance the polariton coherence time. In the last few years many examples of superconducting GHz and THz metamaterials have been presented, both using BCS superconductors like Nb and NbN or high $T_c$ superconductors[20]. For our cavity we choose Nb, which is a well-known superconducting material widely employed in THz science. Superconducting metasurfaces were fabricated by ebeam lithography and etching on top of the same n=1 quantum well 2DEG already used for the measurements reported in Fig. 4(c,d), The Nb film has a critical temperature Tc of 8.7 K, as results from DC resistance measurements[10] . From this value we can infer a gap value of $E_g=2\Delta=4.1$ $K_B$ Tc=3 meV which corresponds to 730 GHz. In Fig.5 (a,b) we report the transmission as a function of temperature for a superconducting metasurface fabricated on semi-insulating GaAs substrate [10]. The design is very similar to what discussed in the previous paragraph and in the superconducting case the resonant frequency is 470 GHz. In this way we can characterize the resonator's response only due to the superconducting material. The effect of the superconducting transition is evident both in the time trace as well as in the Fourier transform of the TDS signal and is well reproduced by our 3D finite element simulation were we use a surface impedance model for the Nb surface [20].

Now we consider the complete system and we analyze the measurements on a second sample where the superconducting metasurface is strongly coupled to the Landau levels of the 2DEG. In Fig.5(c) we report a color plot of a transmission experiment carried out at 2.9 K. A clear anticrossing between upper and lower branch is observed and we can measure a normalized coupling ratio 0.27 which proves that the system is operating in the ultrastrong light-matter coupling regime. Due to the use of a complementary metasurface the polaritonic branches are especially clear because the cyclotron signal gets filtered out. A transmission maximum parallel to the cyclotron dispersion is observable starting from 1.4 T and a frequency of 0.5 THz and is due to the coupling of the second mode of the resonator to the cyclotron resonance. We can model then the complete system with finite element 3D simulations, where, for the cyclotron resonance we use a magnetic field dependent complex conductivity[10]. As visible from the color plot, the experimental data is well reproduced by the numerical modeling. The transmission peaks deduced from the simulations have been reported on the experimental data as black circles.

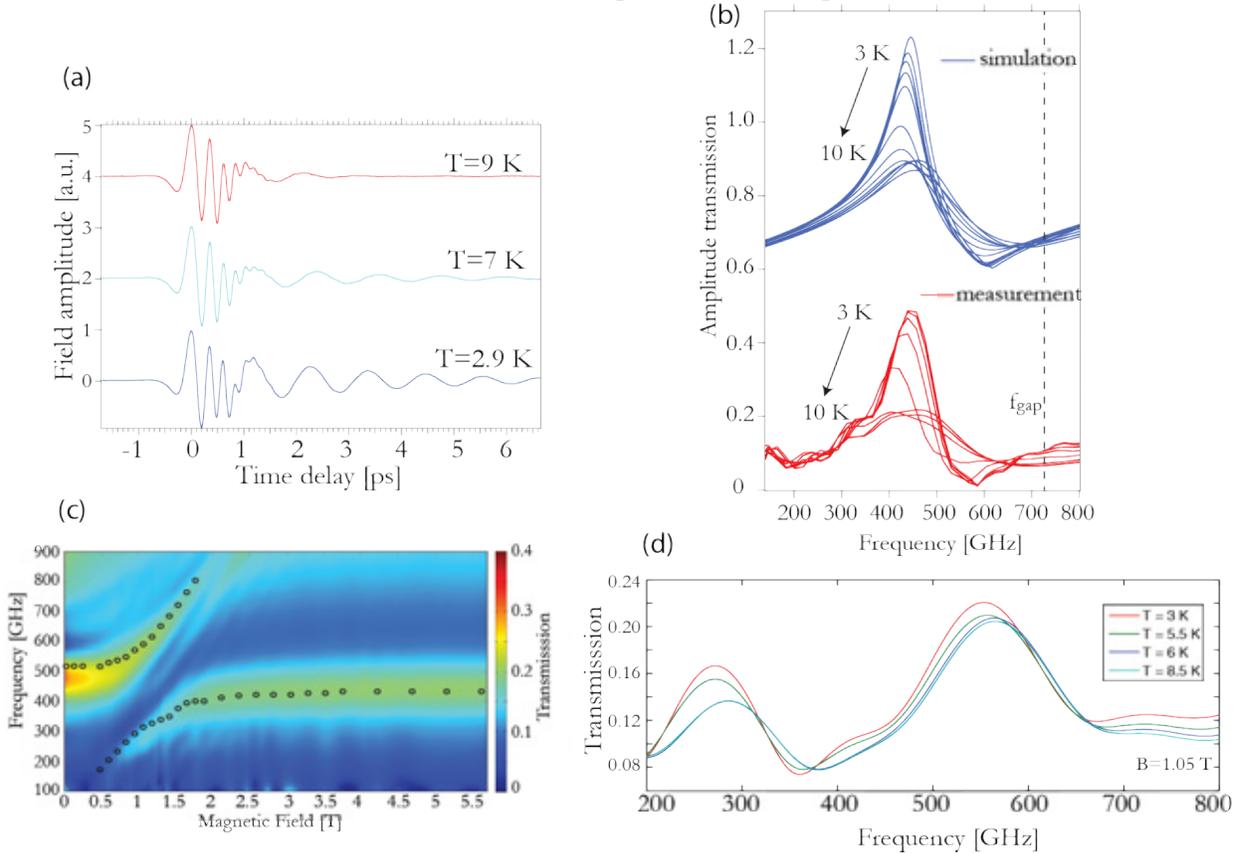

**Fig. 5** : *(a): electric field magnitude as a function of time of the THz-TDS signal transmitted through the metasurface deposited on SI GaAs for three different temperatures. (b): simulated (top) and measured (bottom) transmission for the superconducting metasurface deposited on the SI GaAs as a function of the temperature varied in the range 2.9-10 K. The simulated curves are offset by 0.6 for clarity. (c): transmission color plot for the Nb metasurface deposited on top of the triangular well 2DEG as a function of the applied magnetic field at T=2.9 K.*

*The black circles are the transmission maxima extracted from the numerical simulation. (d): Sample transmission at the anticrossing field of 1.05 T as a function of temperature. Shift in resonance frequency and amplitude transmission is visible on both branches. (Adapted with permission from [10]).*

The possibility to modulate the ultrastrong coupling regime by changing the characteristics of the cavity is demonstrated in Fig.5(d) where we show a series of transmission spectra taken at the anti-crossing field B=1.05 T by changing the sample's temperature from below to above $T_c$. We can clearly observe a modulation both in intensity and in frequency of both polariton peaks.
 For the application of these structures in the context of the non-adiabatic cavity QED experiments [6], we need to enhance this switching capability and then we need to engineer a more suited resonator.  To maximize the switching effect allowed by the presence of the superconductor, we adopted the following design strategy: we increased the radiative Q factor in order to have a structure whose resonance line width is loss limited. The radiative quality factor can be engineered by acting on the capacitor gap dimension in order to reduce the efficiency of the dipolar coupling  as well as the inter-meta-atom spacing . An SEM picture of the fabricated metasurface is reported in Fig. 6 (a) together with 3D simulations of the surface currents. When designing the structure we used the information on the regions of the resonator were the highest current density is flowing which will give rise to the higher ohmic losses. We engineered such regions in order to maximize the effect of the superconducting transition. By employing very narrow (700 nm) and long (13 μm) inductive elements we obtain a very high resistance when the superconductor is above $T_c$. On the contrary, when the Nb is operated below $T_c$ such regions present an higher inductance due to the kinetic inductance of the Cooper pairs and their loss is reduced by the very low, close to zero, value of the real part of the surface impedance.
The high radiative quality factor in this kind of electronic split ring resonator is the result of our optimised design for a superconductor. We chose the symmetric, electronic-like split ring resonator approach in order to control the radiative coupling mainly with the capacitor gap and its width. Due to its symmetry, the coupling to the resonator through magnetic field is negligible and the coupling happens at first order through the capacitor. By adopting a narrow (1 μm gap) capacitor we can increase the radiative quality factor and target a fairly low frequency using long inductors. The use of long inductive elements with a standard metal would be penalizing because of the high ohmic loss, but in our case we employ a superconductor so we keep these losses at the minimum.  The relation which expresses the total quality factor for a resonator as a function of material loss and radiation loss is:

$$\frac{1}{Q_{tot}} = \frac{1}{Q_{rad}} + \frac{1}{Q_{loss}}$$

We can then use the simulated value of the quality factor for the perfect electric conductor (PEC) to deduce the loss quality factor for the measured Nb and the simulated  Au. The Q factor obtained for a PEC is $Q_{PEC}=62= Q_{rad}$ and corresponds to the radiative coupling of the structure, since the material losses are set to zero.  This allows to deduce the loss Q factor for the fabricated Nb resonator and for a simulated structure made with gold. The resonator we propose, if fabricated with gold, presents a very low loss quality factor Q=4 due to the narrow and long wires that result highly dissipative even when fabricated with a good conductor.  On the contrary, the Nb resonator displays a high loss quality factor   $Q_{loss}$=418 when in the superconducting state due to the low value of the real part of the surface impedance and the high value of the reactance. The kinetic inductance in the superconducting phase contributes in a major way to the determination of the resonator's frequency. If we examine Fig.6(d)  we see that for a perfect electric conductor with no kinetic inductance the resonant frequency is 408 GHz which is 1.5 times of what obtained with the Nb.

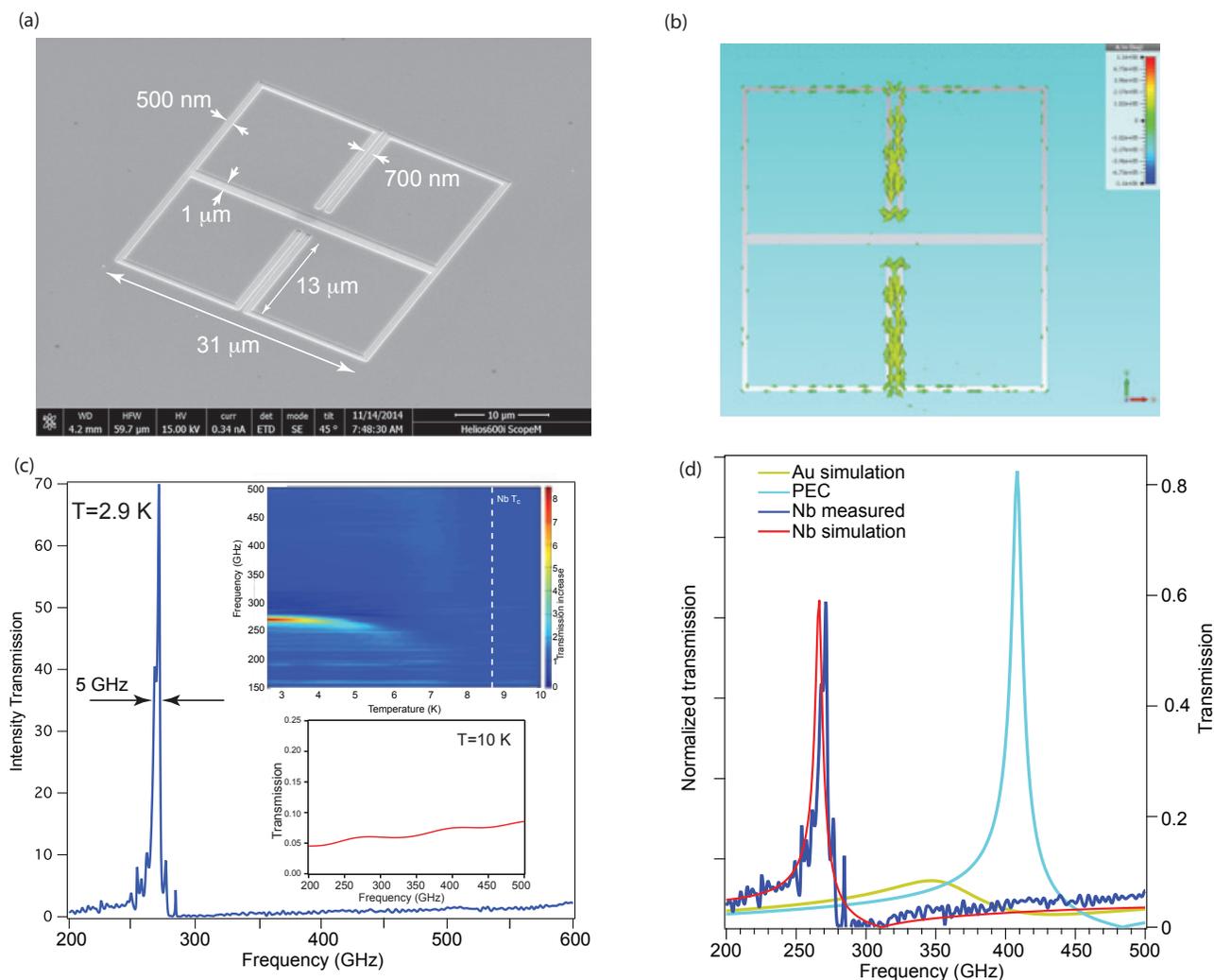

**Fig. 6** *(a): SEM picture of one meta-atom constituting the complementary Nb metasurrface (100 nm thick film ). (b): finite element 3D simulation of the currents flowing in the meta-atom for the Nb in the superconducting state. The highest current density is in the long and narrow wires.(c): Metasurface intensity transmission at T=2.6 K normalised to the transmission value at T=10 K for a long scan of 40 mm. Inset (high): color plot of the metasurface amplitude transmission as a function of the temperature, normalised to the transmission at 10 K. Inset (low): Metasurface transmission at 10 K referenced to empty sample holder: no resonance is observed (d): Experimental transmission at 2.6 K normalized to transmission at 10 K for the Nb metasurface (blue line, left y-axis) together with simulated curve for the same quantity (red line, left y-axis). Simulated transmission for Perfect Electric Conductor (light blue line, right y-axis) and Gold (yellow line, right y-axis). (Reproduced with permission from [21] . Copyright 2014, AIP Publishing LLC).*

We already pointed out that we chose inductances with a very high aspect ratio in order to maximize the resistance when the Nb is in normal state. The same aspect ratio enhances the role of the kinetic inductance in determining the resonator frequency which shifts from the value of 408 GHz for PEC to the measured 269 GHz for Nb [21].

## 4 Signatures of the ultrastrong coupling regime

By exploiting the scaling of the normalized coupling ratio discussed above with both filling factor and number of quantum wells we achieved a record-high coupling of $\frac{\Omega}{\omega_c} = 0.89$ in a sample containing n=4 quantum wells coupled to a Nb superconducting metasurface resonating at 310 GHz. The color plot relative to this experiment is visible in Fig. 7(a). In this case the filling factor for the 2DEG at the anticrossing is about 25 .

In the standard treatment of strongly coupled systems the counter rotating and diamagnetic terms are neglected. The resulting Hamiltonian conserves the excitation number of the coupled system, the resulting splitting is symmetric, and the total energy of the system is unchanged compared to the uncoupled constituents. In contrast, in the ultrastrong coupling limit, different excitation manifolds are mixed, leading to energy shifts in the overall energy[3]. Thereby, the diamagnetic terms counteract the counter-rotating terms. The counter-rotating terms lead to correlations which reduce the energy of the ground state, and in the Dicke Hamiltonian, lead to a phase transition. However, the diamagnetic term leads to a self-interaction of the confined light with the polarization induced in the electric transition by the light itself. It contributes with a positive energy term to the ground state of the coupled system and inhibits the phase transition. At zero magnetic field, the self-interaction due to the diamagnetic term leads to a blue-shift. This dependence is plotted in Fig.7(b), together with the measured data points (the lower point for sample F is omitted since it is outside the bandwidth of the spectrometer). The deviation from the linear case, without counter-rotating and diamagnetic terms (straight gray lines), is evident. The conservation of the total energy in the case of the strong coupling regime is reflected in the symmetry of the gray lines around 1. In contrast, the overall energy of the ultrastrongly coupled system is larger than the summed energy of the single uncoupled systems.

At zero magnetic field, the self-interaction due to the diamagnetic term leads to a blue-shift of the upper polariton, as is visible from the calculation of the polariton dispersion as a function of magnetic field for different coupling strength reported in Fig. 7(c). For high values of the magnetic field, the lower polariton branch for any coupling strength approaches the cold cavity value $\omega_{LC}$. A *polaritonic gap* of width $\Delta\omega$ is created and with a little algebra starting from the polariton branch expression [9] is possible to give an expression which relates this gap to the coupling strength [10,11]:

$$\frac{\Delta\omega}{\omega_{LC}} = \sqrt{\left(\frac{2\Omega}{\omega_{LC}}\right)^2 + 1} - 1$$

Experimental data are plotted against this relation for all the measured samples in Fig.7(d) and there is a very good agreement with the theoretical prediction. The opening of a polaritonic gap is a feature of the ultrastrong coupling regime [7,22,23]. In quantum wells, the origin of it was attributed to the plasma frequency arising from Coulomb interactions of the intersubband transitions[24], overlooking the diamagnetic terms arising from the light-matter interaction. Our results demonstrate that the nonlinear bending of the polariton dispersion is due only to the competition between the diamagnetic terms and the counter-rotating terms of light-matter interaction. Moreover, the diamagnetic term leads to the increase of energy which opens the polaritonic gap[11].

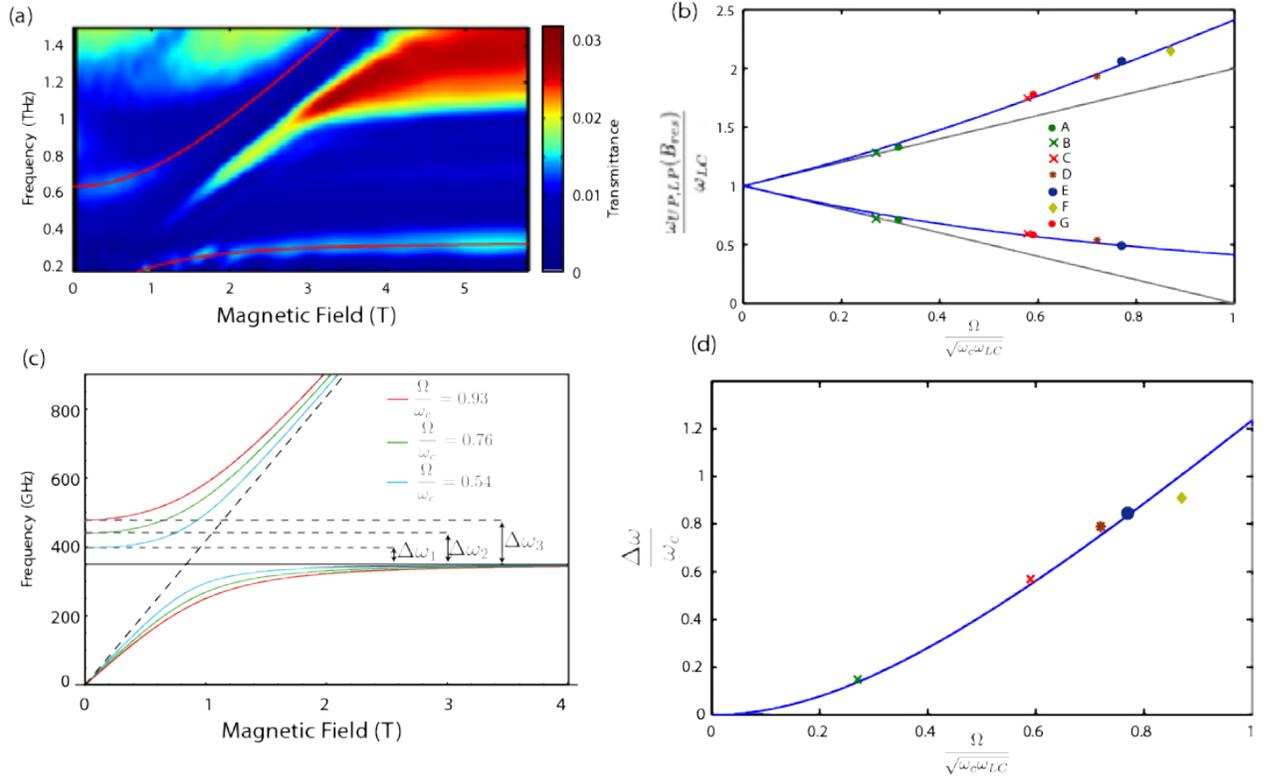

**Fig. 7** *(a): Transmission through sample with Nb resonator with the LC-mode at 310 GHz. (b): The evolution of the normalized polariton frequencies at resonance follows clearly the blue shifted prediction (in blue). The gray lines show the linear behavior expected in the strong coupling regime (without diamagnetic and counter-rotating terms). (c): Calculated polariton branches as a function of the coupling strength for a resonator frequency of 350 GHz. (d): Normalized polaritonic gap as function of the normalized coupling ratio.(Adapted with permission from [11]).*

## 5 Conclusions and perspectives

In this paper we reviewed [10,11,16,17,21] a series of experiments aimed at the study of the ultrastrong light-matter coupling regime in solid-state semiconductor systems at THz frequencies. We presented a new platform for the experimental study of this regime which is based on high-mobility 2DEGs and metasurfaces composed of subwavelength split-ring resonators. We demonstrated record-high normalized splitting ratios of $\frac{\Omega}{\omega_c} = 0.89$ and the intrinsic scalability of the system will allow in the future to reach normalized coupling ratios higher than unity. The study and fabrication of superconducting metasurfaces has also been presented and with such metasurfaces we demonstrated ultrastrong coupling to 2DEGs. We presented a further evolution of the superconducting design which displays high quality factor and holds promises to be proficiently used in non-adiabatic cavity QED experiments[6]. The presented experimental scheme offers a number of interesting opportunities for future experiments involving also cavity-controlled transport, graphene ultrastrong coupling and Dicke physics in solid state [25].

*Acknowledgments*

We gratefully acknowledge financial support from the Swiss National Science Foundation (SNF) through the National Centre of Competence in Research Quantum Science and Technology (QSIT) and through the SNF Grant No. 129823 and by the EU commission through the ERC Advanced Grant *Quantum Metamaterials in the Ultra Strong Coupling Regime* (MUSiC). We would like to acknowledge the support by the FIRST clean room collaborators.

**Giacomo Scalari**

Giacomo Scalari got his physics degree from the University of Pisa, he spent a period in Scuola Superiore S.Anna and he got his PhD from the Université de Neuchâtel, Switzerland. From 2007 he was Post-Doc and then Oberassistant at ETH Zürich. In 2011 he was hired as Permanent Senior Scientits at ETH Zürich in the group of Jérôme Faist. His research activity embraced terahertz quantum cascade lasers, their properties in strong magnetic fields and their extension to low frequencies. He worked on metallic-based microcavity lasers for sub-wavelength emitters, on third-order distributed feedback lasers and on photonic crystal THz quantum cascade lasers. During the last years, he extended his research interests to strong light-matter coupling at THz frequency, time-domain THz magneto-spectroscopy of nanostructures, superconducting metamaterials and quantum-cascade-based frequency combs in the THz range. He was awarded the Swiss Physical Society national prize in 2006 for Applied Physics.